# A Workflow for Correlative In-situ Nanochip Liquid Cell Transmission Electron Microscopy and Atom Probe Tomography Enabled by Cryogenic Plasma Focused Ion Beam

Neil Mulcahy[1], James O. Douglas[1], Syeda Ramin Jannat[1], Lukas Worch[1], Geri Topore[1], Baptiste Gault[1,2], Mary P. Ryan[1], Michele Shelly Conroy[1*]

1. Department of Materials and London Centre for Nanotechnology, Imperial College London, Exhibition Road, London SW7 2AZ, U.K.

2. Max Planck Institute for Sustainable Materials, Max-Planck-Str. 1, 40237 Düsseldorf, Germany

*Corresponding author: mconroy@imperial.ac.uk

## Abstract

Operando/in-situ liquid cell transmission electron microscopy (LCTEM) allows for real time imaging of dynamic nanoscale liquid-based processes. However, due to the thick liquid cell of traditional LCTEM holders and thus scattering of the electron beam passing through the cell, the achievable spatial and chemical resolution is limited. Cryogenic atom probe tomography (cryo-APT) overcomes these limitations by offering (near-)atomic scale compositional analysis of frozen liquid-solid interfaces. However, APT provides limited structural analysis and has no capacity for dynamic or operando liquid cell studies. This work presents a novel workflow for site-specific cryo-APT sample preparation of liquid-solid interfaces from in-situ electrochemical LCTEM Micro-Electro-Mechanical Systems (MEMS) chips. Using cryogenic inert gas transfer suitcase and a cryogenic plasma-focused ion beam (PFIB), a MEMs nanochip containing a Li electrolyte from an electrochemistry LCTEM holder was successfully frozen, transferred to the cryo stage of a PFIB and prepared into APT needle samples containing the electrolyte-electrode interface at cryogenic temperatures, followed by cryogenic transfer to an atom probe for nanoscale compositional analysis. This correlative approach provides dynamic nanoscale imaging and near atomic scale compositional analysis of liquid-solid interfaces. This method enables reliable and reproducible APT sample preparation of these frozen interfaces from MEMs based nanochips and can hence be used across materials systems and energy-conversion or storage devices.

## Introduction

Transmission electron microscopy (TEM), electron spectroscopy and diffraction are invaluable tools in materials science research for investigating the structural, chemical and compositional properties of a variety of different material systems and processes[1]. While traditional TEM techniques can provide up to atomic scale insights into a variety of properties[2, 3], the majority of characterization is done in a static vacuum state. In reality, many of the materials and interfaces require analysis during use of the material for its functional application and thus under dynamic real-world environments[4-6]. Typically, TEM experiments require high vacuum conditions[7], thus restricting the majority of liquid and gas sample analysis[8]. One way the TEM research community has overcome the vacuum restriction is by encapsulating samples of interest in cells with thin electron beam transparent viewing windows[9]. Micro-electro-mechanical systems (MEMS)-based technology has transformed the field of in-situ TEM by facilitating not only the ability to encapsulate the sample in a gas or liquid environment, but also to directly heat or bias the samples in the chips. In combination with new in-situ holder designs, MEMS-based TEM allows for the precise control of the sample environment and applied stimulus while conducting imaging, diffraction and/or spectroscopy[10-13]. One can now probe an extensive list of material phase transformation such as degradation and growth mechanisms in battery materials[14-16] and catalysts[17, 18], and other functional materials in real time at high spatial resolution.

One area which has grown in popularity over the last decade due to this MEMs chip revolution, is the field of in-situ/operando liquid cell TEM (LCTEM). LCTEM allows for real time observations of solution phase static and dynamic processes at high temporal and spatial resolutions[8]. This is achieved through the use of a closed cell environment based on MEMS chip design. The liquid is allowed to flow through the system of interest using external liquid supply systems or syringe pumps[19, 20] while being viewed through electron transparent silicon nitride membranes ($SiN_x$) on both bottom and top. Novel MEMS chip designs have enabled electrical contacts and heating elements to be fitted to the bottom nanochip allowing in-situ/operando electrical biasing and heating experiments within liquid environments[14, 21]. This type of design is also used for in-situ gas closed cell experiments that require heating or biasing[22-26]. A schematic representation of an example of a nanocell can be seen in Figure 1. This type of experimental set up has provided insights into a number of dynamic nanoscale phenomena such as nanoparticle synthesis and growth[27, 28], corrosion[29, 30] and electrochemical phenomena such as dendrite growth [14, 31-33] and solid-electrolyte interphase (SEI) formation[15, 34].

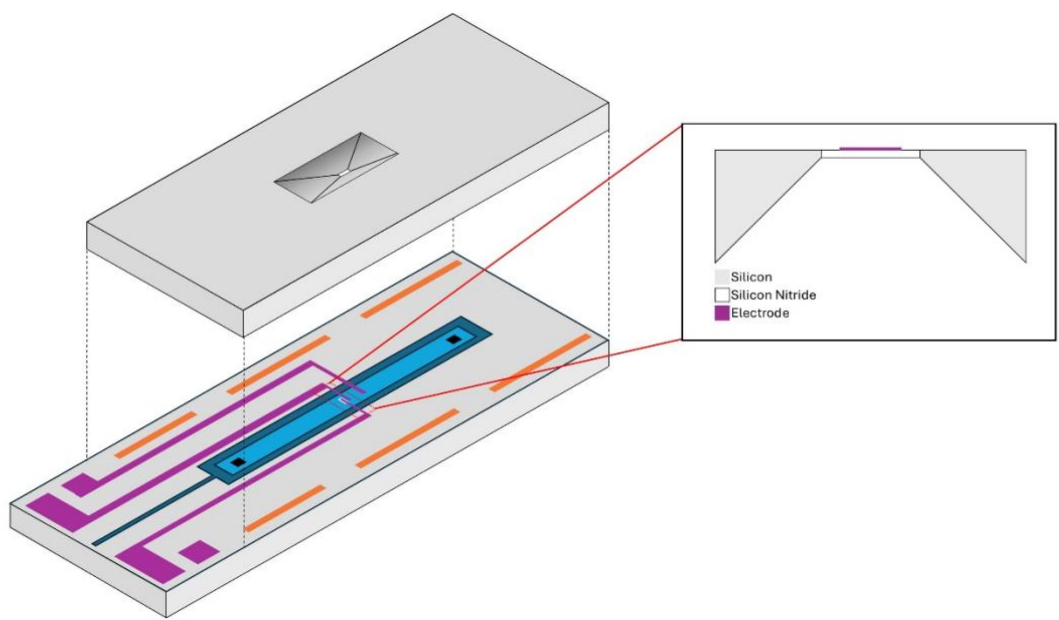

**Fig.1:** A schematic highlighting a liquid cell electrochemical nanocell, composed of a top and bottom MEMs based Si nanochip. The $SiN_x$ window with printed electrode is highlighted.

While the usefulness of this microscopy technique for investigating liquid processes at the nanoscale has been detailed in numerous reports, the technique still presents key problems particularly with respect to reduced spatial resolution[35] and electron beam induced effects[36, 37]. Reduced spatial resolution in LCTEM in comparison to traditional TEM/(S)TEM largely occurs due to increased electron scattering as a result of the increased sample volumes. The overall cell thickness typically comprises of the liquid, the electron transparent membrane windows and the sample of interest. Often this can lead to an overall cell thickness in excess of 100 nm, making it difficult for electrons to be transmitted through the cell. This means imaging and spectroscopy techniques such as electron diffraction or electron energy loss spectroscopy (EELS) are either at significantly lower resolutions or cannot be performed at all. LCTEM studies are therefore typically only capable of generating dynamic nanoscale imaging and the only available way of determining any compositional changes is through the generated contrast, be it from phase- or amplitude-contrast. While various solutions have been proposed and implemented, such as altering cell designs and using 2D window cells[15, 38-40] and utilising dose limitation techniques, many of these solutions remain frontier and challenging to apply to commercially available liquid cell systems and nanochips. Initial publications to combine closed liquid cell with biasing have allowed for near atomic resolution, however liquid flow is not possible in these set-ups [41, 42].

The ability to provide both dynamic nanoscale imaging and nanoscale compositional analysis within liquid environments would be invaluable for problems affecting numerous material systems, particularly within battery research. However, any such correlative approach would require the liquid environment to remain in its state of interest. In recent years, cryogenic microscopy techniques such as cryogenic TEM/STEM and cryogenic atom probe tomography (cryo-APT)[43-47] have been used to provide near atomic scale compositional insights into various liquids and liquid-solid interfaces[46, 48]. Such cryogenic analysis has been realised through advances in site specific specimen preparation workflows by using cryogenic-focused ion beam/scanning electron microscopy (FIB/SEM),[49-53]. While cryo-APT uniquely combines chemical sensitive, high resolution 3D compositional mapping, it has no capacity for dynamic or operando studies and only provides snapshots of the evolution of a particular system, with limited crystallographic/structural analysis. This makes information provided by in-situ/operando LCTEM and cryo APT extremely complementary to one another.

In this work, we present a novel correlative workflow which provides site-specific cryo-APT specimen preparation of a liquid solid-interface from a liquid cell electrochemical MEMs-based nanochip. By integrating cryogenic inert gas transfer technology and a cryogenic plasma-FIB/SEM (PFIB/SEM), we have successfully frozen a MEMs chip covered in a lithium electrolyte from a commercial LCTEM holder, transferred the frozen interface to the cryo stage of a PFIB/SEM and prepared reliable and reproducible cryo-APT specimens from the frozen liquid-solid interface. The created cryogenic samples were transferred directly to the analysis chamber of an atom probe under cryogenic conditions. We captured the 3D nanoscale compositional analysis of the frozen liquid-solid interface from the MEMs nanochip.

## Instrumentation and Materials

### Liquid cell Transmission Electron Microscopy system and MEMs chips

A Thermofischer Scientific (Waltham, Massachusetts, United states) Spectra 300 (S)TEM at 300 kV accelerating voltage was used for all STEM measurements. This instrument is probe corrected and fitted with an ultra-high-resolution X-FEG Ulti-monochromator. The measured screen current during imaging was 47 pA, which equates to an electron dose of $1.22 \times 10^3$ e/Å$^2$. High angle annular dark field (HAADF) imaging was used for all (S)TEM images shown.

The Stream system, including the in situ liquid TEM holder and the pressure based liquid supply system (LSS), supplied by DENSsolutions B.V. (Delft, The Netherlands) was used for all work involving LCTEM. A nanocell is comprised of a top and bottom silicon wafer chip. The MEMs based bottom nanochip contains a three-electrode set up, with a Pt reference, counter and working electrode. The working electrode is deposited on a 50 nm electron transparent $SiN_x$ membrane window. This window has dimensions of approximately 20 μm x 200 μm[21]. The top chip contains an identical electron transparent membrane window, allowing for viewing of the working electrode within the microscope. Prior to assembly, the bottom nanochip is plasma cleaned in Ar-O for approximately 3 minutes. This will make the nanochip hydrophilic, which is advantageous when trying to freeze the sample to avoid large droplets of liquid. A commercial Lithium electrolyte, $LiPF_6$ in ethylene carbonate/diethyl carbonate (EC/DEC), supplied by Merck Life Science UK Ltd (Dorset, United Kingdom), was flown through an assembled cell within the TEM. The flow was controlled by the LSS in combination with Impulse, a commercially available software from DENSsolutions B.V. (Delft, The Netherlands). Within the software an inlet and outlet pressure of 2000 and -950 mbar respectively were selected, and this will result in a flow rate of ≈ 8 μL/min.

**Vacuum Cryo Transfer Module and Glovebox**

Samples could be transferred under vacuum/inert conditions and also at constant cryogenic temperatures between instruments using a Vacuum Cryo Transfer Module (VCTM), supplied by Ferrovac GmbH (Zürich, Switzerland). The module is equipped with a small ion pump with a non-evaporable getter cartridge, allowing the module to maintain a pressure of $10^{-10}$ mbar. Cryogenic temperatures can be maintained within the module using a dewar of liquid nitrogen (LN$_2$). The module can accept industry standard pucks and cryo pucks supplied by CAMECA Inc. (Gennevilliers, France). The system contains a 500 mm wobblestick with a PEEK-insulated puck manipulator allowing pucks to be picked up or released[51].

All cryogenic sample preparation was conducted using an inert glovebox, supplied by Sylatech Ltd. (York, United Kingdom). This glovebox contains an inert nitrogen atmosphere with typical oxygen content and humidity both below 5 ppm during operation. It possesses a large load lock, allowing for large samples or even entire TEM holders to be inserted directly into the glovebox chamber. LN$_2$ can be pumped directly into a bath within the glovebox using a large pressured dewar external to the system, Apollo 50, supplied by Cryotherm Inc. (Kirchen (Sieg), Germany). Samples can be plunge frozen within the glovebox and transferred to a VCTM using a combination of the LN$_2$ bath and a cooled "elevator" contained within a loadlock chamber capable of being pumped to vacuum. This loadlock chamber is connected to a Ferroloader docking station, supplied by Ferrovac GmbH (Zürich, Switzerland). Frozen samples can be picked up directly from the elevator loadlock chamber using a wobblestick within the VCTM and pulled into the module, maintaining the sample under vacuum and at constant cryogenic temperatures throughout the entire process.

**Cryogenic Plasma Focused Ion Beam/Scanning Electron Microscope**

A Helios Hydra CX (5CX) plasma FIB from Thermo Fisher Scientific (Waltham, Massachusetts, United states) fitted with an Easylift tungsten cryo-micromanipulator and an Aquilos cryo-stage was used for all FIB/SEM work shown. Through the circulation of gaseous nitrogen passing through a heat exchanger within a large external dewar of LN$_2$, the stage and micromanipulator could be cooled to approximately 90 K. To achieve this base temperature, a nitrogen gas flow of 180 mg/s was maintained. The temperature of both the stage and micromanipulator could be controlled using a temperature control unit, Model 335 cryogenic temperature controller supplied by LakeShore Cryotronics Inc (Westerville, Ohio, United States) in combination with heaters built into the stage. The system is also equipped with a Ferroloader docking station. This allows a precooled VCTM to be docked to the side of the instrument and cryogenic samples under vacuum can be inserted directly to the cryo stage of the FIB from the module without heating up[51].

For the purpose of this work, a "Dual-puck" holder stage baseplate supplied by Oxford Atomic (Oxford, United Kingdom) was used. This allows for two industry standard cryo pucks supplied by CAMECA Instrument Inc. (Madison, WI, USA), to be inserted to the cryo stage at once. The typical configuration used for these experiments was one puck containing the frozen liquid cell nanochip, while the other puck would contain a pre-prepared Si microarray coupon. The Si microarray coupon was prepared at room temperature prior to any cryogenic work. This involves pre-cutting the posts at 0°, while also precoating the posts and micromanipulator in SEMGlu™, supplied by Kleindiek Nanotechnik GmbH (Reutlingen, Germany) [54, 55]. This pre-preparation procedure is detailed by Mulcahy et al.[56]. The FIB column is set at 52° with respect to the SEM column. Any FIB work shown used Xenon plasma.

**Atom Probe Tomography**

A Local Electrode Atom Probe 5000 XR supplied by CAMECA Instruments Inc. (Madison, WI, USA) was used for all atom probe analysis shown. This instrument is equipped with a reflectron system and a Ferroloader docking system. This means a VCTM can be docked directly to the atom probe and a specimen can be inserted onto the cryo stage located in the atom probes analysis chamber through the use of a "piggyback" puck while being maintained under vacuum and at cryogenic temperatures. The sample was analysed using laser pulsing analysis (40-80 pJ, 80-240 kHz, 1 ion per 500 pulses on average, 25k base temperature). 3D reconstructions and atom probe data analysis were completed using AP suite 6, a commercially available software from CAMECA Instruments Inc. (Madison, WI, USA.).

# Results and discussion

### Liquid cell TEM

Figure 2 shows HAADF STEM images of (a) an electrode prior to any liquid flow and (b) after Li electrolyte has flown into a nanocell. The change in imaging resolution and quality is apparent following liquid flow. This decreased resolution is largely occurring as a result of increased electron beam scattering due to the thickness of the liquid within the nanocell. For the purpose of this demonstration no electrical bias was applied. The liquid was allowed to flow through the system for ten minutes to replicate a liquid-solid interface that would form during an operando biasing experiment to serve as a model system to showcase this correlative workflow.

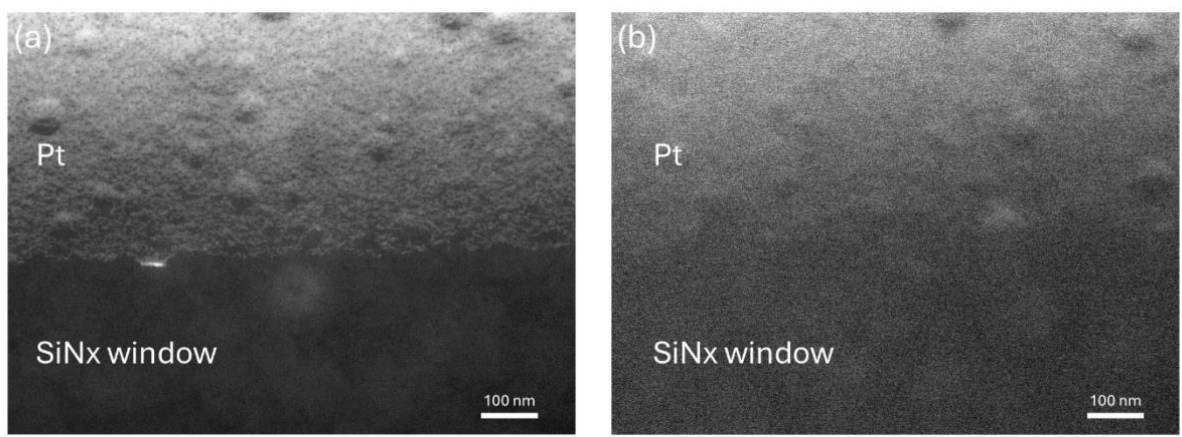

**Fig. 2:** HAADF STEM images of a Pt electrode during a LCTEM experiment (a) prior to any liquid flow and (b) as Li-containing electrolyte is flowing through the system

### Freezing process

Following the conclusion of the LCTEM experiment, both inlet and outlet valves on the holder are closed, ensuring liquid remains within the system and free from direct air contact, and all connections from the holder to the liquid supply system (LSS) are completely removed. Simultaneously, a copper block (in this instance the "Dual-puck" holder stage baseplate supplied by Oxford Atomic) was placed in a small bath of $LN_2$ and allowed to cool to $LN_2$ temperatures ($\approx 77.15$ K), making it act as a "cold block". The LCTEM holder is removed completely from the TEM and disassembled in air. Upon disassembling the bottom nanochip contained a noticeable layer of liquid electrolyte over the electrodes/$SiN_x$ membrane window. This bottom chip is subsequently placed on the cold block to rapidly cool the system to $LN_2$ temperatures, preserving the liquid-solid interface frozen in its state of interest. Following noticeable freezing of the liquid on the chip, the entire chip would then be completely submerged in $LN_2$ to protect it from frost build up. Figure 3 shows a schematic and photo of this freezing process, highlighting the setup used.

While Li-containing electrolytes, electrodes and Li decomposition products are air- and moisture - sensitive[57-59], it was deemed that speed was a more important factor when freezing the bottom nanochip, and that immediately disassembling the nanocell in air rather than transferring to the inert nitrogen glovebox and subsequent freezing would lead to better preservation of any grown decomposition products (such as SEI components), and any degradation of the region of interest would be protected to some extent by the volume of the electrolyte covering it. Further to this, previous experiments by the author had shown that directly plunge freezing the nanochip in LN$_2$ resulted in the loss of decomposition products. Slower cooling on a cold block was found to work better for maintaining the region of interest of the sample. This method would lead to increased frost build up on the chip while cooling but this

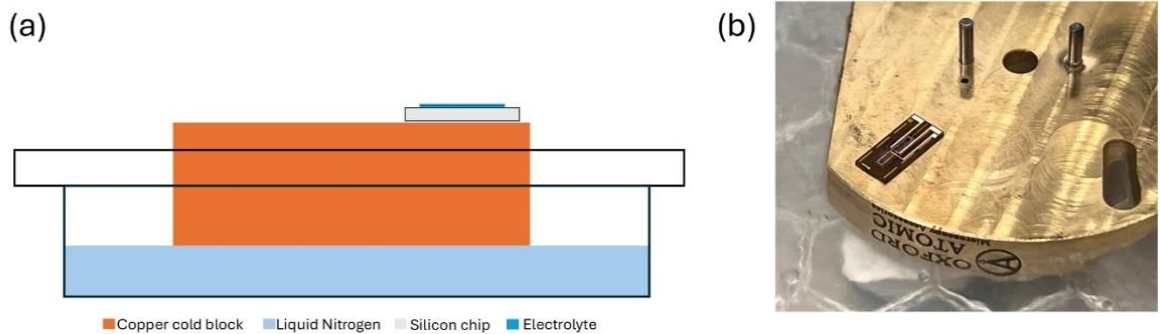

**Fig.3:** (a) A schematic highlighting the freezing process where a Cu block submerged in LN$_2$ with an electrolyte covered nanochip placed on top can be seen. (b) A photo showing this freezing process, with a noticeable layer of frozen electrolyte on top of the LC nanochip.

**Transfer process**

The next step of the workflow involves transferring the now frozen liquid-solid interface on the MEMs nanochip to the cryo stage of a PFIB/SEM. This is achieved through the use of an inert nitrogen glovebox and a VCTM. Figure 4 presents schematics of the step-by-step procedure involved in transferring a frozen sample through the glovebox, into a VCTM and subsequently to the cryo PFIB/SEM. Frozen nanochips can be introduced into the glovebox through a side antechamber using a small volume of LN$_2$, illustrated in Figure 4 (a). This process will evaporate a small amount of LN$_2$ during the pumping and purging of the antechamber. However, it was found to be sufficient to maintain the sample in its frozen state. Within the glovebox a larger bath can be filled with LN$_2$ from an external pressurised dewar. The frozen nanochip is inserted into this larger bath, ensuring the ROI remains frozen. This process is shown in Figure 4 (b). From here the frozen nanochip is placed into a cryo clip on a cryo puck, which is compatible with the VCTM, the stage baseplate of the cryo PFIB/SEM, and the APT instrument. A loadlock within the glovebox can be cooled using the external pressurised dewar, and the frozen nanochip on the cryo puck can be inserted into this loadlock through the use of an elevator, as seen in Figure 4 (c). The elevator loadlock can be maintained at approximately 115 K. This loadlock can be closed and subsequently pumped to UHV. The vacuum level and temperature within this loadlock are sufficient to prevent frost build up, while maintaining the sample in its frozen state. From here the sample can be pulled into a precooled VCTM, which once filled with LN$_2$ can maintain a temperature of approximately 90 K, as shown in Figure 4 (d). The VCTM can be detached from the glovebox, loaded onto the PFIB/SEM using a Ferroloader docking station, and the frozen sample can be inserted directly to the cryo stage of the PFIB/SEM, without heating up or building up frost.

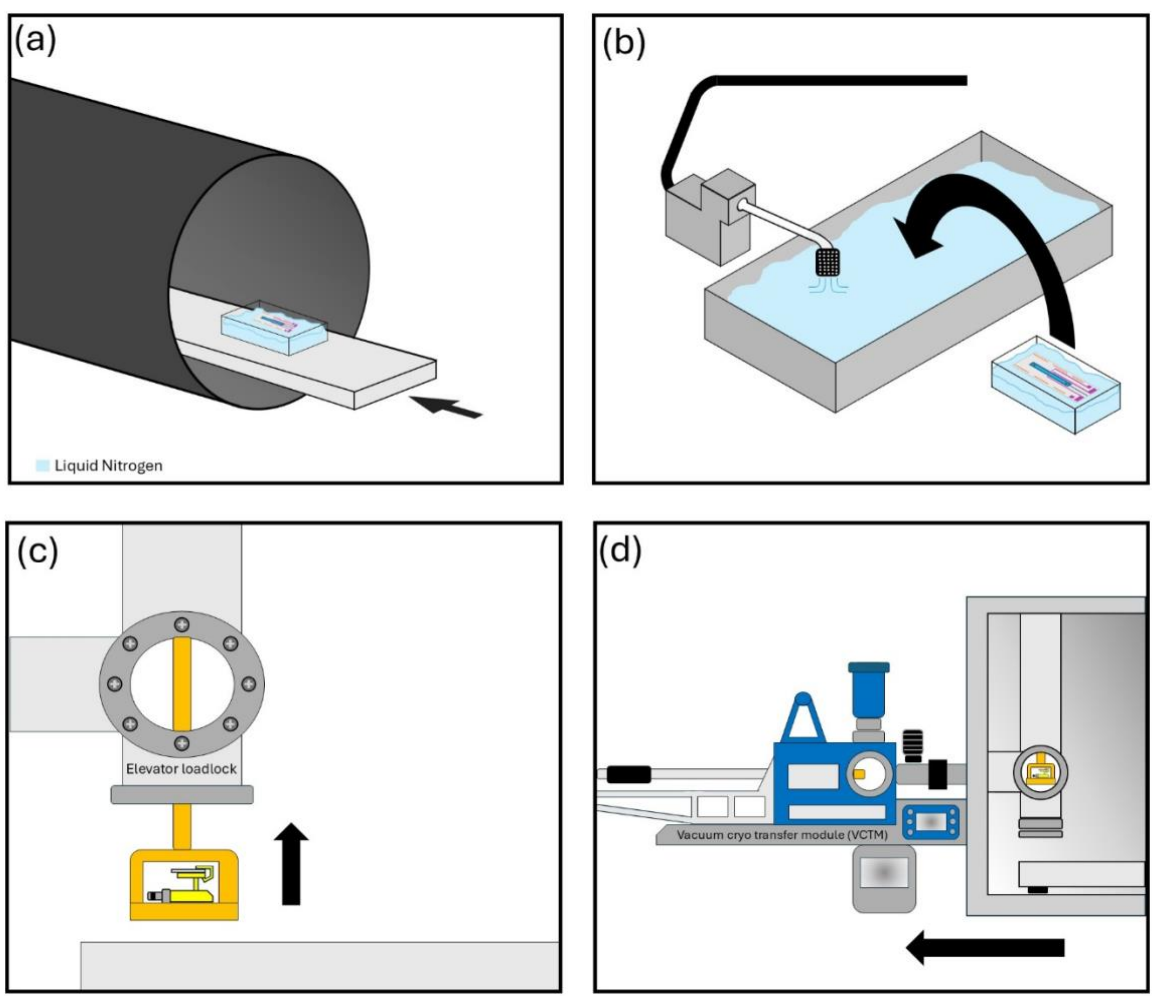

**Fig. 4:** Schematics highlighting the transfer process for moving a frozen MEMs nanochip to the cryo stage of a PFIB/SEM through the use of an inert glovebox and a VCTM. (a) shows the frozen nanochip being inserted into the glovebox through a small side antechamber using a small volume of $LN_2$. (b) shows the frozen nanochip being moved from this small volume to a larger bath of $LN_2$ within the glovebox, maintaining the system in its state of interest. (c) the frozen nanochip is then placed into a cryo clip on a cryo puck and inserted into the elevator within the glovebox. This elevator can be raised into a loadlock chamber and pumped to UHV. (d) the frozen nanochip under UHV can be transferred directly to a precooled VCTM.

**Cryogenic FIB/SEM preparation**

The aim for this part of the workflow is to create a specimen appropriate for APT analysis from the frozen liquid-solid interface involving the Pt electrode and Li electrolyte. APT specimen requirements involve creating a needle-like geometry with a diameter of approximately 100 nm or below at the apex[48]. The first step for this process involves identifying the ROI using the SEM beam. Figure 5 (a) and (b) showcase two separate examples of a Pt electrode covered in electrolyte. The uniformity of the electrolyte on the electrode can vary substantially between experiments, often making it difficult to identify the exact location of the ROI. This non-uniformity can be attributed to various parameters such as the flow rate of liquid during the in-situ experiment, the freezing process, and whether the nanochip has been plasma cleaned prior to insertion into the liquid cell holder. It is recommended to use a higher kV electron beam (20-25 kV) to directly see the electrodes through the electrolyte. If even at higher kVs the electrodes cannot be identified due to the thickness of the electrolyte, milling around the ROI may be required. Due to the thin nature of the $SiN_x$ membrane window and electrode directly milling will quickly leave a hole which is easy to identify in comparison to milling a bulk part of the nanochip.

Once the ROI has been identified it is necessary to add a protection layer prior to any further milling to protect the interface from damage from the ion beam. At room temperature this protection layer is typically achieved through the use of a decomposed organometallic gas injected via a Gas injection System (GIS), forming site specific layers of various metals such as Pt or W, or other species such as C[60]. However, at cryogenic temperatures this site-specific deposition is more challenging to control as the precursor gas will condensate over mm² sized areas of the cold sample surface[61, 62], ultimately losing the site specificity of the process. This can make it exceptionally challenging to identify the ROI. While possibilities of controlling this process have been detailed by Parmenters et al.[63], a more reliable site-specific method is required. As described by Schwarz et al.[52] and for flat surfaces by Woods et al.[50], site specific in-situ deposition of a metallic layer can be achieved by rastering the ion beam over a lamella attached to the micromanipulator that is in close proximity to the sample's surface. This process is demonstrated in Figure 5 (d) and (e), where the Xe ion beam (30 kV, 1 nA) is being rastered over the surface of a Cr lamella, coating the surface of the electrolyte-electrode in a thin site-specific protective layer of Cr.

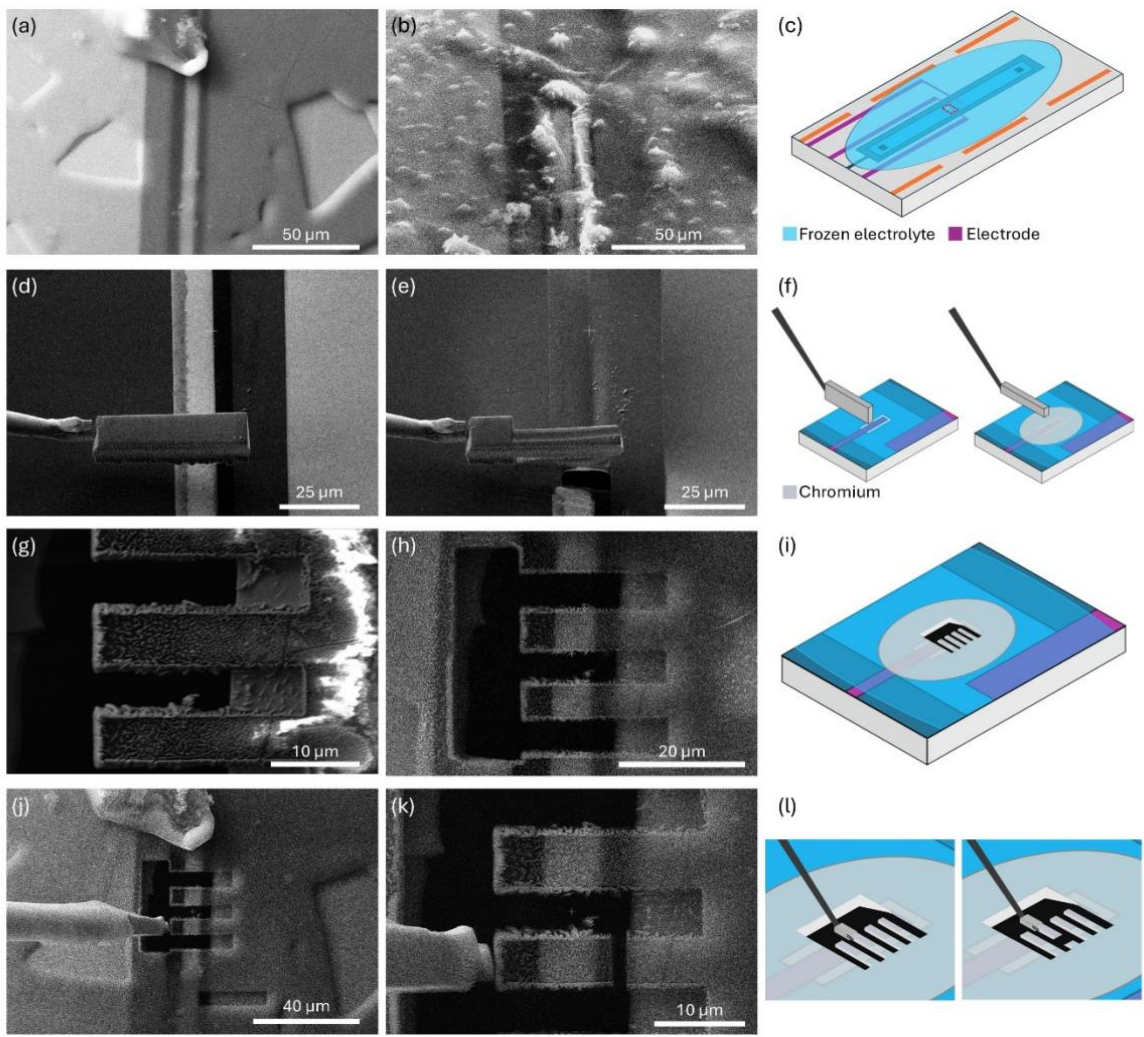

**Fig. 5:** SEM micrographs and schematics highlighting the ROI identification, Cr protective layer coating, milling and lift out procedure. (a) and (b) show two examples of an electrode covered in electrolyte, with (c) highlighting the area of the MEMs nanochip which is being observed. (d) and (e)

show how a Cr protective layer is applied over this ROI, with (f) showing a schematic representation of what is occurring during this process. (g) and (h) show the ROI milled into liftout bars. (g) was taken at 5 kV and (h) at 25 kV, highlighting how a higher kV SEM beam is effective for identifying the position of the electrode. (i) shows a schematic of the liftout bars relative to the overall MEMs nanochip. (j) shows how redeposition welding with a Cr lamella was used to attach a liftout bar to the micromanipulator and in (k) this sample is milled free from the nanochip and lifted out. This lift out process is schematically shown in (l).

Ensuring a sufficient protective layer is covering the ROI, the FIB stage is tilted to 52° and using the Xe plasma beam (30kV, 1-4 nA) the electrodes of interest were milled into liftout bars, taking care not to damage the ROI at the apex of the electrodes. An example of this can be seen in Figure 5 (g) & (h). Due to the thin nature of the $SiN_x$ membrane window and electrode an undercut is not performed, and attempting to create an undercut could potentially result in damaging the ROI. Again, due to the difficulties involved in using the GIS at cryogenic temperatures the created liftout bars could be lifted out using a preprepared Cr lamella using redeposition welding (Xe, 30 kV, 30 pA), Figure 5 (j) & (k), as described in[49] and[50]. Schematics of the ROI identification, Cr coating, milling and lift out procedure can be seen in Figure 5 (c), (f), (i), and (l).

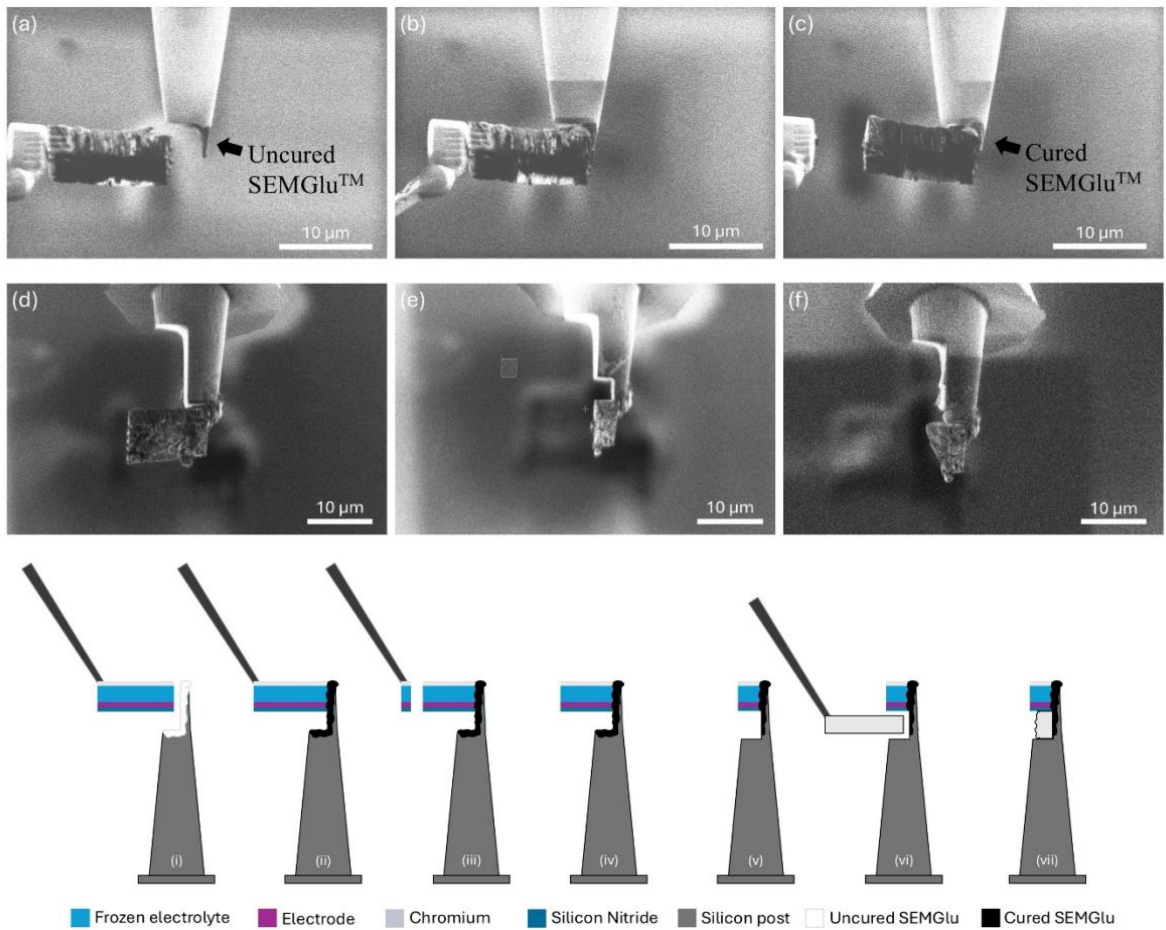

**Fig. 6:** (a)-(c) present SEM micrographs of the liftout bars being attached to Si microarray posts using SEMGlu™. Note the contrast change of the SEMGlu™ on the post from (a) to (c) due to the glue being cured. (d) to (e) shows how the sample has been cut to fit the post, and also how the interface between the liftout bar and post has been altered to allow a better connection to be made. (f) shows the same sample that has been filled in with Cr using redeposition. This entire procedure is shown schematically from (i) to (vii).

Due to the difficulty with creating an undercut when lifting out the electrode-electrolyte interface it can be challenging to securely attach a sample to a Si microarray post at cryogenic temperatures using redeposition welding alone, due to the beam sensitive nature of the sample, and the lack of good surface contact between the post and the sample. As showcased in Figure 6 (a-c), the liftout bar can be brought in contact with frozen SEMGlu[TM] on a preprepared Si microarray post. This glue can be cured using the ion beam (Xe, 30 kV, 30 pA), and the sample will remain securely stuck to the post once milled free from the micromanipulator without further assistance[56]. Note the changing contrast of the glue once cured. Following this the sample is milled to fit the dimensions of the Si post, as seen in Figure 6 (d), and the connection between the post and the sample is also milled, Figure 6 (e), to allow for a more secure contact to be created using redeposition from a Cr lamella, as described by[49]. This is done to ensure the sample is mechanically stable and electrically conductive throughout its entire volume. An example of a filled in sample can be seen in Figure 6 (f). This entire procedure is schematically shown step by step in Figure 6 (i-vii).

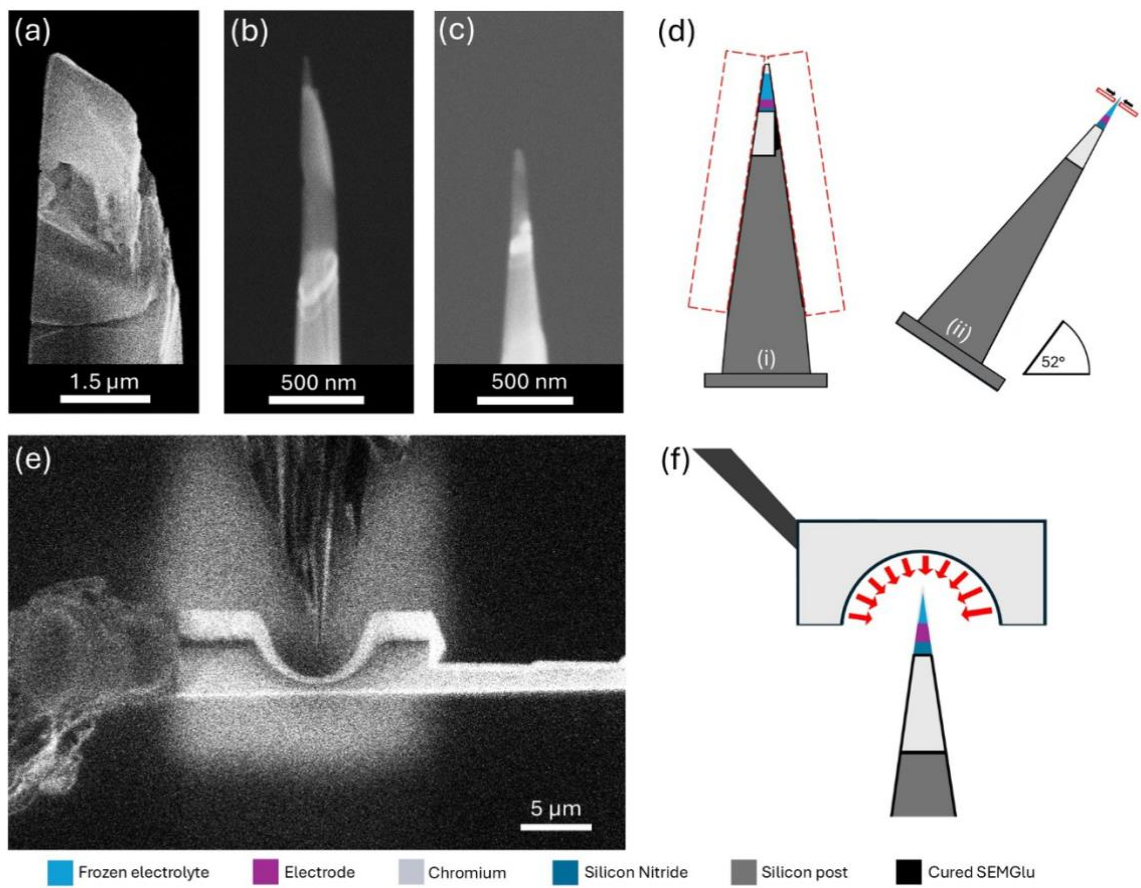

**Fig. 7:** SEM micrographs showing a sample after (a) rough milling at 0° stage tilt, (b) fine milling at 52° stage tilt, and (c) following a low kV ion beam polish. (d) showcases a schematic of both the rough and fine milling procedures. (e) SEM micrograph of the coating process with (f) representing this schematically.

At this point the sample is ready to be milled into a needle shape with the aim of creating a specimen with a diameter of roughly 100 nm at the apex. Initially a rough needle shape is created at shallower and shallower shank angles at 0° stage tilt and rotating periodically using a series of rectangular box milling patterns with the ion beam (Xe, 30 kV, 1-4 nA) until a rough diameter of around 1 μm is achieved. The sample is then rotated to 52° stage tilt and again using a series of rectangular box milling patterns with the ion beam (Xe, 30 kV, 10 pA – 0.3 nA) the sample is thinned to be roughly 200 nm in diameter. The difference in sample diameter is highlighted in Figure 7 (a) and (b). A final low kV Xenon plasma beam polish was used to create the final needle shape which can be seen in Figure 7 (c). Schematics of both milling processes are shown in (d). Varying layers are evident in this final needle including electrolyte, Pt electrode and Cr weld indicating that an interface has been captured. As a final preparation step the needle was coated in Cr using redeposition from the ion beam (Xe, 30 pA, 30 kV) on all four sides for 20 s. This process has been described in detail in[52, 64, 65]. Metallic coatings have been shown to offer an effective shield against in-situ delithiation effects that can occur during APT analysis due to the applied electrostatic fields [64]. The coating has also been shown to provide increased mechanical support and superior pathways for heat dissipation. An SEM micrograph of this process and a schematic can be seen in Figure 7 (e) and (f). Created frozen needles containing the electrolyte-electrode interface could then be transferred directly to the analysis chamber of the APT instrument using a pre-cooled VCTM, maintaining the samples in their frozen state and under vacuum throughout the entire process.

## Cryogenic APT

The APT analysis generated a mass spectrum and detection hit map which can be found in detail in Supplementary Figure S1. Two specific ranges have been selected based on peaks containing species of interest that would be expected from the electrode and electrolyte and are shown in Figure 8 (a) & (b). The electrolyte used for this experiment was $LiPF_6$ in EC/DMC ($C_3H_4O_3$/OC(OCH3)2), meaning species containing variations of Li, P, F, O, C and H were assigned as being part of the electrolyte. While residual H will always be present within the atom probe chamber and can obscure H atoms that have been detected from the created specimen[66]. It is thus difficult to distinguish between the two sources of H ions. For this work all detected H species have been assigned as "electrolyte species" in the analysis. From Figure 8 (a) it can be seen that Pt could be readily identified, along with peaks pertaining to PtH and in (b) various species from the electrolyte can be identified such as Li, C and H.

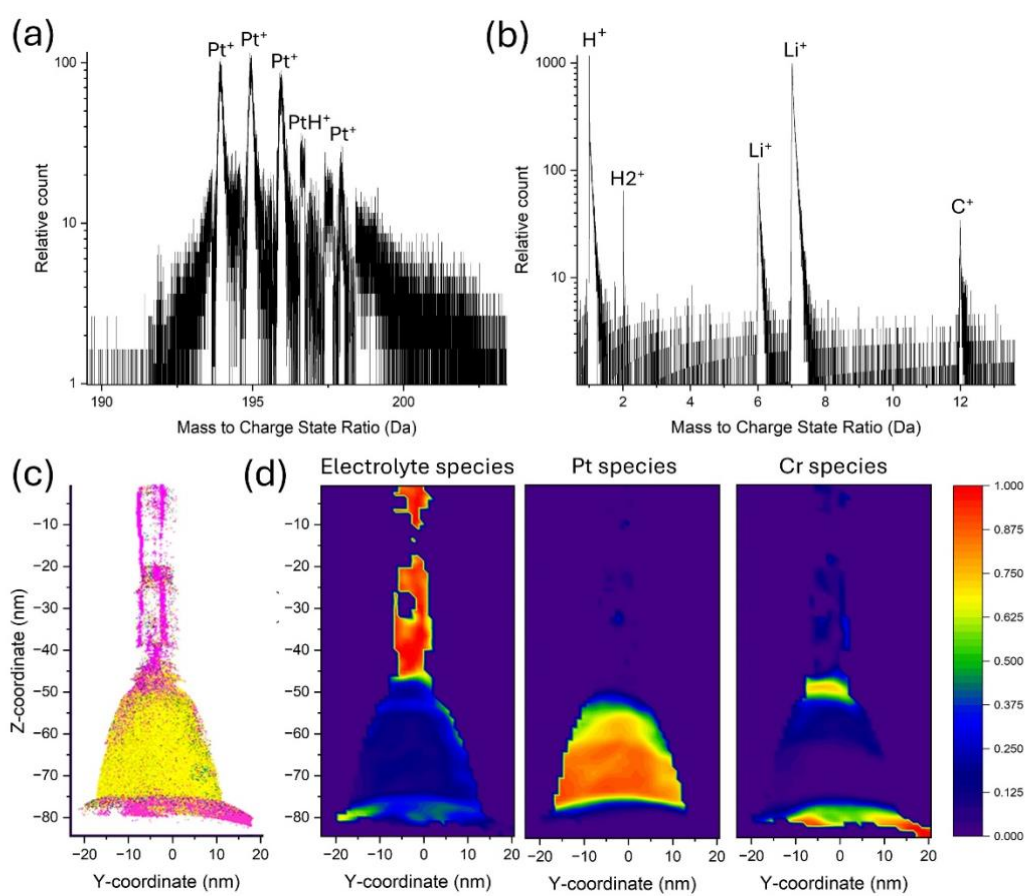

**Fig. 8**: (a) displays a portion of the generated mass spectrum from the APT analysis capturing detected Pt species, while (b) displays another portion of the mass spectrum capturing various electrolyte species such as Li, C and H. (c) shows a 3D reconstruction of the created needle specimen with (d) displays 2D contour plots showing the relative decomposed concentration of electrolyte, Pt and Cr species.

Through the combination of the generated mass spectrum and detection hit map, a 3D reconstruction was generated giving a clear picture of the positions of the detected ions within the created needle specimen and is shown in Figure 8 (c). To give a clearer picture of the ions within the 3D reconstruction 2D contour plots showing the relative concentration of decomposed electrolyte species, Pt species and Cr species is shown in Figure 8 (d). Decomposed 2D contour plots of Pt, Cr and all electrolyte species in X-Z and Y-Z planes can be found in supplementary Figure S2 and S3. Based on the reconstruction and 2D contour plots it is evident three distinct regions have been captured. On the bottom a large concentration of Cr has been detected, translating to the Cr "fill in" from the sample preparation process. Following this there is a large concentration of Pt species, capturing the Pt electrode from the nanochip and above this there is a large concentration of "electrolyte species" from the electrolyte and a smaller concentration of Cr from the Cr coating process at the end of the sample preparation. These distinct regions showcase that the interface between the electrolyte and electrode from the LCTEM has been captured to some extent. Evidently the 3D reconstruction of the region containing the electrolyte species appears to have evaporated non uniformly, with some delithiation effects evident. While a Cr coating was applied and can be seen in the 2D contour plot, it was not sufficient to stop Li migration during analysis. This could of course be due to a non-uniform coating or the running conditions used but this could also point to a need for new methods to reduce Li migration from a frozen electrolyte. Details on all electrolyte and Cr detected species are represented in supplementary figure S4 and S5 as bar charts displaying atom type versus total count.

## Conclusion and outlook

This work provides a novel and reproducible workflow for site-specific cryogenic APT sample preparation of liquid-solid interfaces from MEMs based in-situ nanocells. Through the integration of cryogenic inert gas transfer, a cryogenic PFIB/SEM and an inert nitrogen glovebox with $LN_2$ capabilities, we have successfully preserved, extracted and analysed a Li electrolyte – Pt electrode interface from an in-situ LCTEM electrochemical nanocell using cryogenic APT, maintaining the sample in its state of interest throughout the entire process. The developed workflow overcomes the limitations of traditional LCTEM by enabling sub nanometre three-dimensional compositional analysis, complementing the operando capabilities and dynamic nanoscale imaging offered by LCTEM.

Our findings demonstrate the feasibility of using cryogenic APT in combination with other microscopy techniques, to take advantage of the critical chemical and compositional information APT can offer at (near-)atomic length scales, which is not accessible due to resolution constraints in many liquid-based microscopy techniques. This cryogenic workflow can be adapted to a wide range of liquid based in-situ MEMs studies including various battery materials, catalytic reactions, and corrosion studies across a wide range of techniques such as LCTEM, liquid cell atomic force microscopy, liquid cell synchrotron X-ray imaging etc.

Challenges remain for analysing frozen liquid-solid interfaces using cryo APT, demonstrated by the Li migration effects within the data shown. Further research is required to minimise these effects and provide more representative analysis. This workflow opens up the field of correlative operando/in-situ microscopy and cryogenic techniques. The integration of cryogenic multimodal approaches, including cryogenic electron microscopy and cryogenic APT, has the potential to revolutionise our understanding of solid-liquid interactions at the nanoscale.

## Acknowledgements

N.M., M.P.R., M.S.C. acknowledge funding from Engineering and Physical Sciences Research Council (EPSRC) and Shell for funding through the InFUSE Prosperity Partnership (EP/V038044/1). This work was made possible by the EPSRC Cryo-Enabled Multi-microscopy for Nanoscale Analysis in the Engineering and Physical Sciences EP/V007661/1. M.S.C. acknowledges funding from Royal Society Tata University Research Fellowship (URF\R1\201318) and Royal Society Enhancement Award RF\ERE\210200EM1. L.W. and G.T. EPSRC Centre for Doctoral Training in the Advanced Characterisation of Materials (CDT-ACM)(EP/S023259/1) for funding their Ph.D. studentships. And G.T. acknowledges Cameca Ltd. For co-funding their PhD. SRJ thanks PhD funding from the Faraday Institution, under the grant EP/S514901/1. B.G. acknowledge financial support from the ERC-CoG-SHINE-771602.

## Contributions

M.C. conceived the idea of combining liquid cell TEM with site specific cryo APT analysis. N. M., J.O.D., S.R.J., L.W., M.S.C. conducted the SEM and FIB experiments, N.M., and M.C. did the in-situ TEM experiments, N.M., J.O.D., B.G., M.S.C., analysed the APT specimens and processed the data. N.M. and M.S.C. lead the publication writing. All authors discussed the results and contributed to the final version of the manuscript. The project was supervised by B.G., M.P.R. and M.C.

## Corresponding authors

Michele Conroy, mconroy@imperial.ac.uk

## Conflict of Interest

The authors declare that they have no known competing financial interests or personal relationships that could have appeared to influence the work reported in this paper.

# A Workflow for Correlative In-situ Nanochip Liquid Cell Transmission Electron Microscopy and Atom Probe Tomography Enabled by Cryogenic Plasma Focused Ion Beam

Neil Mulcahy[1], James O. Douglas[1], Syeda Ramin Jannat[1], Lukas Worch[1], Geri Topore[1], Baptiste Gault[1,2], Mary P. Ryan[1], Michele Shelly Conroy[1*]

1. Department of Materials and London Centre for Nanotechnology, Imperial College London, Exhibition Road, London SW7 2AZ, U.K.

2. Max-Planck Institut für Eisenforschung GmbH, Düsseldorf 40237, Germany

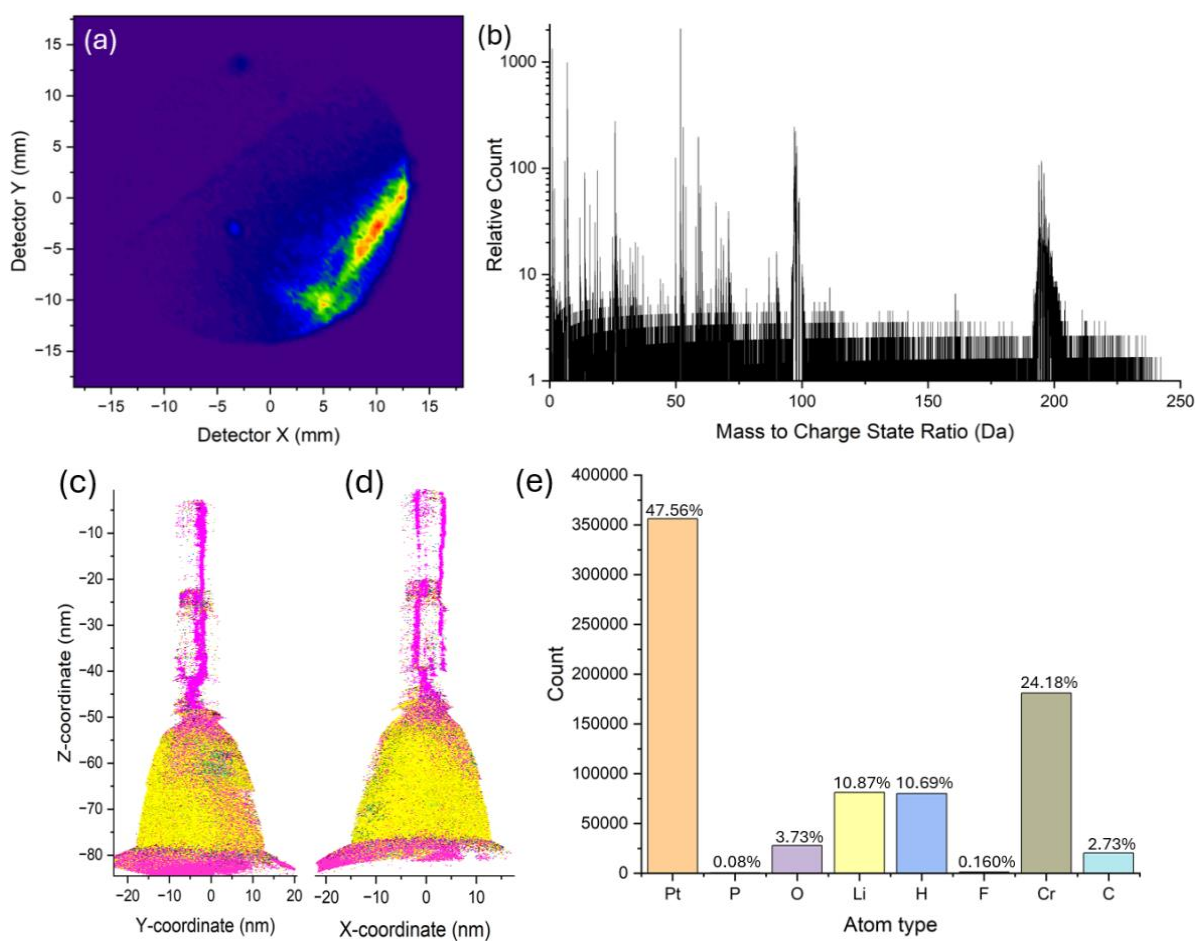

**Fig. S1:** (a) Detection hit map and (b) full mass spectrum from APT analysis of Pt-Li electrolyte interface. (c) 3D reconstructions generated from (a) and (b) in X and Y directions. (e) a bar chart showing total decomposed species count versus atom type, with the percentage of each type of atom present from all the ranged species listed.

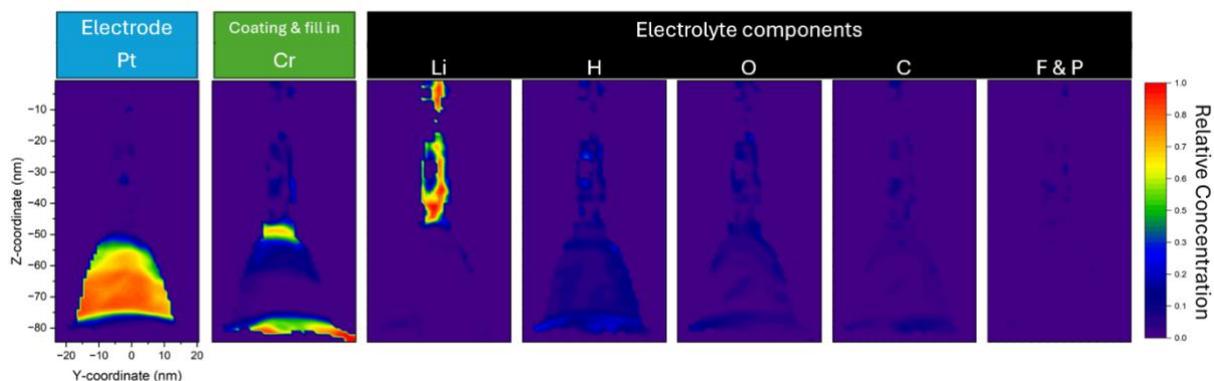

**Fig. S2:** 2D contour plots showing the relative concentrations of Pt, Cr and various electrolyte species including Li, H, O, C , F and P within the 3D reconstruction in the Y-Z plane.

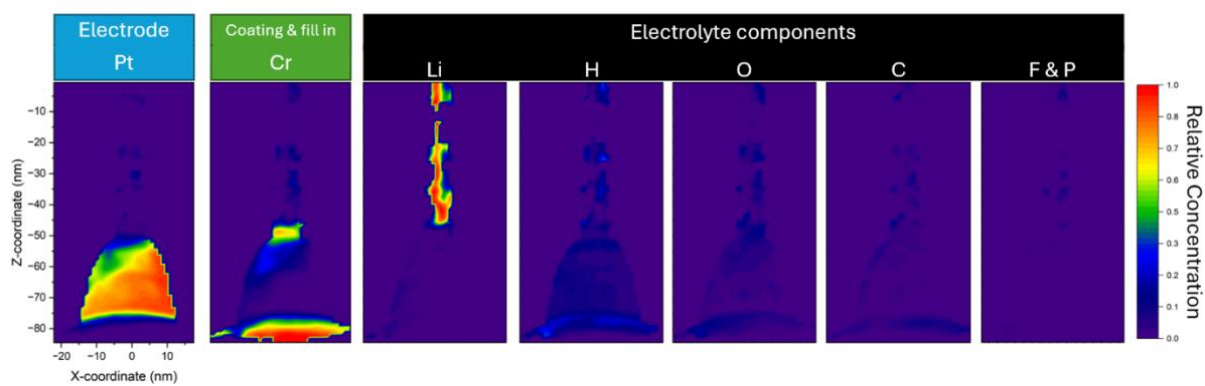

**Fig. S3:** 2D contour plots showing the relative concentrations of Pt, Cr and various electrolyte species including Li, H, O, C , F and P within the 3D reconstruction in the X-Z plane.

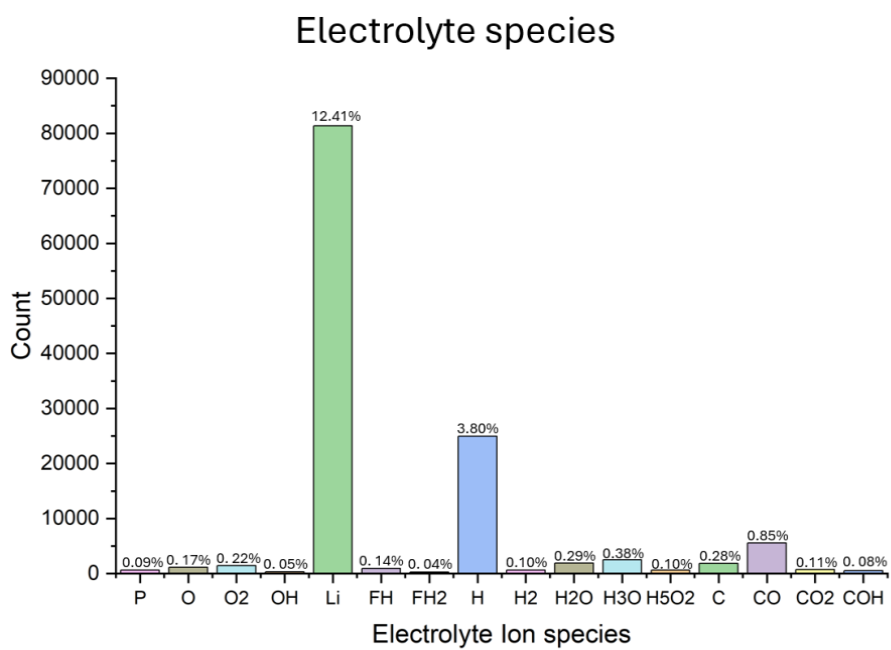

**Fig. S4**: Bar chart showing every ionic species that contained electrolyte species (H,O,P,F,Li) versus detected count. The percentage of detected species versus total count of all detected species is shown over each bar.

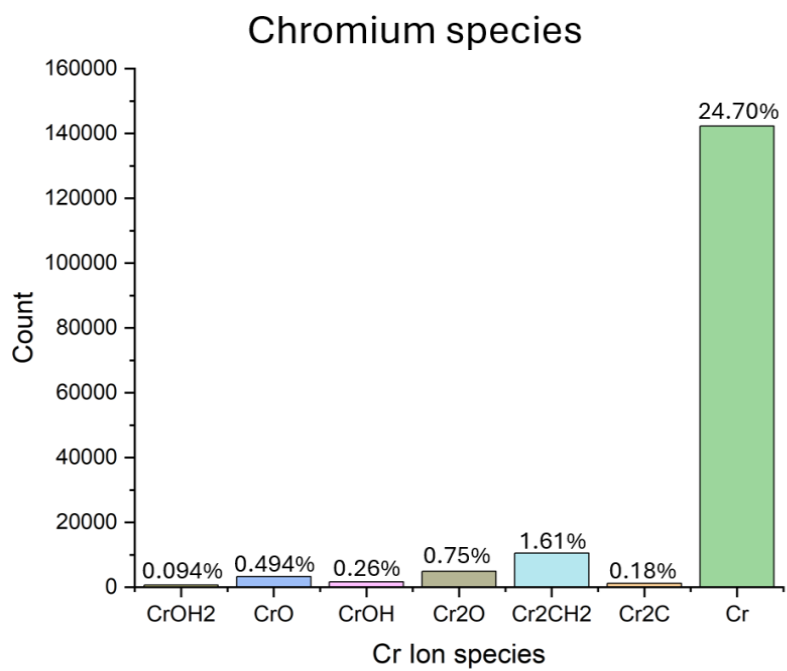

**Fig. S5:** Bar chart showing every ionic species that contained Cr versus detected count. The percentage of detected species versus total count of all detected species is shown over each bar.